\documentclass[aps,pre,12pt]{revtex4-1}

\usepackage{graphicx}
\usepackage{graphics}
\usepackage{color}
\usepackage[intlimits,tbtags]{amsmath}
\usepackage{amsfonts,amssymb,amsthm}

\def\beq{\begin{equation}}
\def\eeq{\end{equation}}
\def\bea{\begin{eqnarray}}
\def\eea{\end{eqnarray}}

\begin{document}

\title{One-dimensional discrete aggregation-fragmentation model}

\author{N. Zh. Bunzarova~$^{\dag^1\dag^2}$,
N. C. Pesheva~$^{\dag^2}$ and J. G. Brankov~$^{\dag^1\dag^2}$  
\email{brankov@imbm.bas.bg}}

\affiliation{$^{\dag^1}$~Bogoliubov Laboratory of Theoretical Physics,
 Joint Institute for Nuclear Research, 141980 Dubna, Russia}

\affiliation{$^{\dag^2}$~Institute of Mechanics, Bulgarian
Academy of Sciences, 1113 Sofia, Bulgaria}

\begin{abstract}

We study here one-dimensional model of aggregation and
fragmentation of clusters of particles obeying the stochastic
discrete-time kinetics of the generalized Totally Asymmetric
Simple  Exclusion Process (gTASEP) on open chains.  Isolated
particles and the first particle of a cluster of particles hop one
site forward with probability $p$; when the first particle of a
cluster hops, the remaining particles of the same cluster may hop
with a modified probability $p_m$, modelling a special kinematic
interaction between neighboring particles, or remain in place with
probability $1-p_m$. The model contains as special cases the TASEP
with parallel update ($p_m =0$) and with sequential
backward-ordered update ($p_m =p$). These cases have been exactly
solved for the stationary states and their properties thoroughly
studied. The limiting case of $p_m =1$, which corresponds to
irreversible aggregation, has been recently studied too. Its phase
diagram in the plane of injection ($\alpha$) and ejection
($\beta$) probabilities was found to have a different topology.

Here we focus on the stationary properties of the gTASEP in the
generic case of attraction $p<p_m<1$ when
aggregation-fragmentation of clusters occurs. We find that the
topology of the phase diagram at $p_m =1$ changes sharply to the
one corresponding to $p_m =p$ as soon as $p_m$ becomes less than
$1$. Then a maximum current phase appears in the square domain
$\alpha_c(p,p_m)\le\alpha\le 1$ and $\beta_c(p,p_m) \le \beta \le
1$, where $\alpha_c(p,p_m)= \beta_c(p,p_m)\equiv \sigma_c(p,p_m)$
are parameter-dependent injection/ejection critical values. The
properties of the phase transitions between the three stationary
phases at $p< p_m <1$ are assessed by computer simulations and
random walk theory.


\end{abstract}

\maketitle

\section{Introduction}

 Different variants of TASEP model are  widely studied currently, since it
 is believed that this model can be helpful in understanding various
 types of systems in Nature, such
as: kinetics of protein synthesis \cite{MGP68,ZLH}, molecular
motors on a single track \cite{TC08}, colloid particles moving in
narrow channels \cite{GGNSC07,K07, ZPB09}, and vehicles on a
single-lane road \cite{N96,CSS00,AFS17,H01}, etc. Another
motivation for studying TASEP-like models is the aim for a better
understanding of the properties of  systems in nonequilibrium
steady states, nonequilibrium phase transitions and various
phenomena with no counterpart in the equilibrium case. In the
model under consideration here, the particles obey the dynamics of
the generalized Totally Asymmetric Simple Exclusion Process
(gTASEP), which essentially is the TASEP with backward-sequential
update equipped with two hopping probabilities: $p$ and $p_m$. The
modified hopping probability $p_m>p$ describes a kinematic
attraction between neighboring particles which hop during the same
integer-time moment. In principle, the model admits the study of
aggregation-fragmentation phenomena, fluctuations and finite-size
effects in nonequilibrium stationary states induced by the
boundary conditions.

We remind the reader that the original TASEP was defined as a
continuous-time Markov process (random-sequential update in the
Monte Carlo simulations) and was solved exactly with the aid of
recurrence relations by Domany et al \cite{DDM92} for special
values of the model parameters, and by Sch\"{u}tz and Domany
\cite{SD93} in the general case. A breakthrough in the methods for
solving TASEP on open chains marks the matrix-product
representation of the steady-state probability distribution, found
in \cite{DEHP}. Different versions of this approach, known as the
Matrix Product Ansatz (MPA), were used also to obtain exact
solutions for the stationary states of TASEP and ASEP under
several types of discrete-time stochastic dynamics:
sublattice-parallel \cite{H96,HP97}, forward-ordered and
backward-ordered sequential \cite{RSS96,RS97}, and fully parallel
(simultaneous updating of all sites) \cite{ERS99}, \cite{dGN99}.
The above studied cases of TASEP     
show that the only dynamics that allow  clusters to move forward
as a whole entity is the backward-ordered one. Then, the
probability for translation of a cluster of $k$ particles one site
to the right is $p^k$, while such a cluster is broken into two
parts with probability $p - p^k$. Under the generalized TASEP
dynamics, the probability for translation of a cluster of $k$
particles one site to the right becomes  $p p_m^{k-1}$, and its
fragmentation into two parts happens with probability $p(1-
p_m^{k-1})$.

We note that the above generalized backward-ordered dynamics was
suggested as exactly solvable one by W\"{o}lki \cite{W05}, and
studied on a ring in \cite{DPPP,DPP15,AB16}. The limit case of
$p_m = 1$ corresponds to irreversible aggregation, or jam
formation in the case of vehicles, suggested and studied by
Bunzarova and Pesheva \cite{BunP17} and further elaborated 
in \cite{BPPB17,BBPP18}. Here, we focus on the stationary
properties of gTASEP when $p<p_m<1$, describing the generic case
of attraction between particles hopping stochastically and
unidirectionally in discrete time along finite one-dimensional
chains with given boundary conditions at the ends.

\section{The model and some known results}
\subsection{The model}

We consider an open one-dimensional lattice of $L$ sites.  An
occupation number $\tau_i$ is associated with a site $i$, where
$\tau_i=0$, if site $i$ is empty and $\tau_i=1$, if site $i$ is
occupied. The dynamics of the model follows the discrete time
backward-sequential rules \cite{RSS96,RS97}. During each discrete
moment of time $t$, an update of the configuration of the whole
chain with $L$ sites, labelled by $i =1,2, \dots, L$, takes place
in $L+1$ steps, passing through successive updates of the right
boundary site $i =L$, all the pairs of nearest-neighbor sites in
the backward order $(L-1,L), \dots, (i,i+1),\dots, (1,2)$, and,
finally, the left boundary site $i=1$ is updated. According to the
generalized backward-sequential rules:

1. Each integer time moment $t$ (configuration update) starts with
the update of the last site of the chain: if site $i=L$ is
occupied, the particle at it is removed from the system with
probability $\beta$ and stays in place with probability $1-\beta$.
If the last particle has left the system, then the particle at
site $i = L-1$ takes its place at site $i=L$ with modified
probability $p_m$; otherwise, it remains immobile with probability
$1-p_m$.

2. Next, a particle at site $i=1,2,\dots L-2$ hops to an empty
site $i+1$ with probability $p$ or $p_m$, depending on the update
of the next nearest neighbor on the right-hand side at the current
moment of time. If site $i$ is occupied and site $i+1$ is empty at
the beginning of the current update, then the particle from site
$i$ jumps to site $i+1$ with probability $p$ and stays immobile
with probability $1-p$. Alternatively, if site $i+1$ is occupied
at the beginning of the current moment of time $t$, and became
empty after the particle from $i+1$ jumped to the empty $i+2$,
then the particle at site $i$ jumps to site $i+1$ with probability
$p_m$ and stays immobile with probability $1- p_m$.

3. The left boundary condition also depends on the occupation
history  of the right nearest-neighbor. If site $i=1$ was empty at
the beginning of the current update, a particle enters the system
with probability $\alpha$ or site $i=1$ remains empty with
probability $1-\alpha$. Alternatively, if site $i=1$ was occupied
at the beginning of the current moment of time, but became empty
under its current update, then a particle enters the chain with
probability $\tilde{\alpha}$ or the site remains empty with
probability $1-{\tilde \alpha}$, where \beq \tilde{\alpha} =
\min\{1, \alpha{p_m}/p\}. \label{lbc} \eeq   We note that the
above left boundary condition was introduced first by Hrab{\'a}k
in \cite{PhD} and, independently, in \cite{BunP17}. It is
necessary for consistency with the special cases of
backward-ordered sequential update, when $p_m=p$ and
$\tilde{\alpha} =\alpha$, as well as with the parallel one, when
$p_m=0$  
and $\tilde{\alpha} =0$.

\subsection{Known results in particular cases}

\begin{figure}[t]
\includegraphics [width=.475\textwidth]{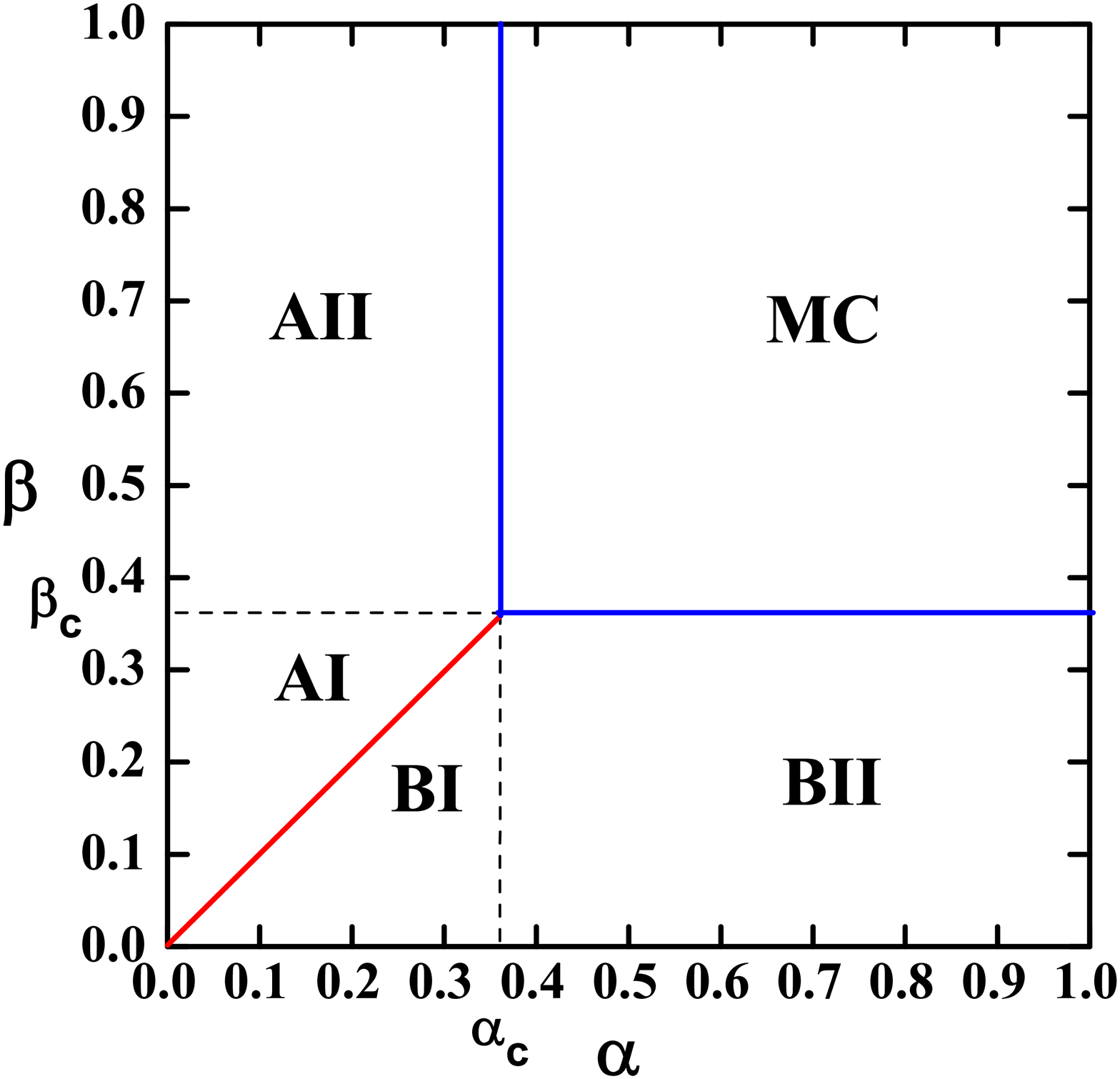} \quad  
\includegraphics [width=.47\textwidth] {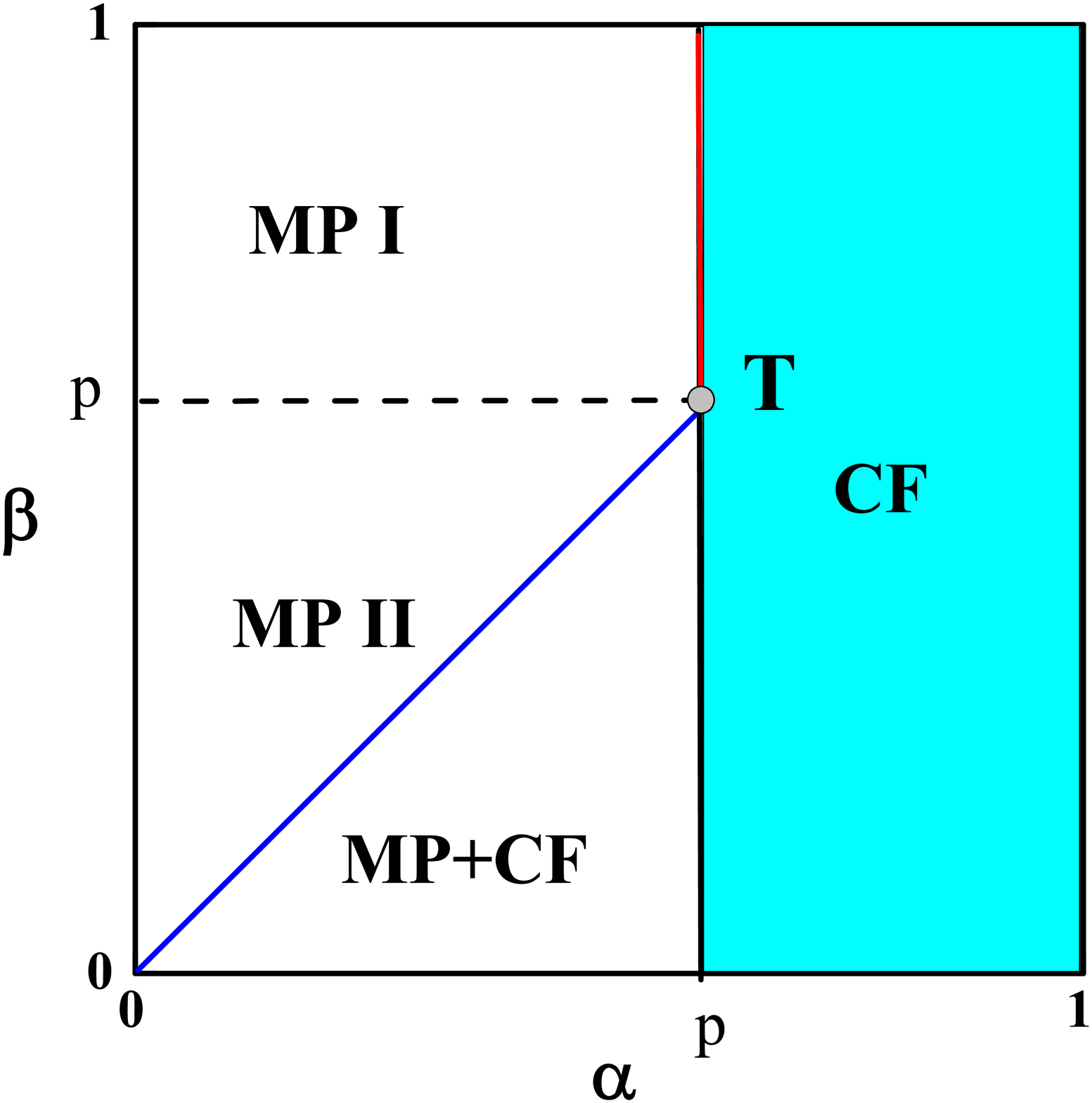}  \\ 
\quad (a) \hspace{.475\textwidth} (b) \caption{ (Color online)
Phase diagrams in the $\{\alpha,\beta\}$-plane of: (a) the
standard TASEP with backward-sequential update with $p=p_m=0.6$.
There are three
stationary phases: maximum-current phase, MC, low-density phase,       
LD=AI$\cup$AII, and high-density phase, HD=BI$\cup$BII; (b) the
gTASEP with $p=0.6$ and $p_m =1$. The three stationary phases are
of different nature: a many-particle phase MP=MPI$\cup$MPII, a CF
phase of completely filled chains, and a  mixed MP+CF phase.}
\label{PhD}
\end{figure}
\medskip
Here we summarize the known results about the phase diagrams and
the  phase transitions between the stationary phases in the
particular cases of $p_m =p$ (the ordinary backward-sequential
update) and $p_m =1$ (irreversible aggregation). The corresponding
phase diagrams are shown in Fig. \ref{PhD}.

In Fig.~\ref{PhD}(a) the subregions AI (BI) and AII (BII) differ                   
by the shape of the local density profiles. The nonequilibrium
phase transitions between low-density phase, LD=AI$\cup$AII, and
maximum-current phase, MC, as well as between high-density phase,
HD=BI$\cup$BII, and MC, are continuous, while the transition
between LD and HD is discontinuous, with a finite jump in the
local density. The exact critical injection/ejection rate values
are $\alpha_c = \beta_c = 1 - \sqrt{1-p}$. In our case of $p=0.6$,
$\alpha_c = \beta_c = 0.367544\dots $.

In Fig.~\ref{PhD}(b) the many-particle phase MP contains a
macroscopic number of particles or clusters of size $O$(1) as
$L\rightarrow \infty$; MPI and MPII differ only by the shape of
the local density profile, which results from the different type
of evolution of the configuration gaps. In the region of MPI, the
inequality $\beta >p$ leads to growing average width of the
rightmost gap, hence the profile bends downward near the chain
length. In the complementary region MPII, the opposite inequality
$\beta < p$ holds and the rightmost gap is short-living, while the
gaps on the left-hand side of it have a critical type of evolution
with mean lifetime of the order $O(L^{1/2})$, see Ref.
\cite{BBPP18}. The phase MP+CF is mixed in the sense that the
completely filled configurations are perturbed by short living
gaps entering the chain from the first site. The configurations of
the stationary nonequilibrium phase CF represent a completely
filled chain with current $J = \beta$. The unusual phase
transition, found in \cite{BunP17}, takes place across the
boundary $\alpha = p$ between the MPI and CF phases.

\section{The generic case of attraction}

Firstly, we aim here to analytically approach the question of how
the completely filled phase (CF) at $p_m=1$ (see Fig.
\ref{PhD}(b)) is destroyed, when $0 < 1- p_m \ll 1$, and
transformed into new phases typical for $p_m < 1$  (see Fig.
\ref{PhD}(a)). One of the methods, developed in Refs.
\cite{BPPB17,BBPP18}, and  which is used also here, is based on
the study of the time evolution of single gaps in different
regions of the CF phase.

Secondly, we present results of extensive computer simulations,
which suggest the topology of the modified phase diagram, the
shift of the triple point $(\alpha_c(p,p_m), \beta_c(p,p_m))$
under the change of $p_m \in [p, 1]$ at fixed $p$, and the nature
of the phase transitions between the stationary nonequilibrium
phases.

\subsection{Time evolution of configuration gaps}

We begin with finding out the probability of a single gap
appearance  under boundary conditions corresponding to the CF
phase. Then we consider the first step in the time evolution of
the gap width. The problem is rather complicated because the
probability of appearance of a gap is position dependent when $p_m
<1$. In contrast to the case of $p_m =1$, here we show that when
$\beta \not= p$, the gap width performs a special, position
dependent random walk.

Let $P_i(p,p_m)$ denote the probability of appearance of a single
gap  at site $i = 1,2,\dots,L$ in a completely filled
configuration with $\tau_1 = \tau_2 = \dots \tau_L =1$, when $0 <
1- p_m \ll 1$. For brevity of notation, we do not show the
explicit dependance on the injection and ejection rates of that
probability. Since we exclude the appearance of a second gap,
under the generalized backward-sequential update we obtain, \beq
P_{L-k}(p,p_m)=(1- p_m)p_m^k\beta, \, k=0,1,\dots L-2,\quad
P_1(p,p_m)= (1- \tilde \alpha)p_m^{L-1}\beta. \label{probgap} \eeq

Under the assumption $0<1- p_m \ll 1$, the left boundary condition
yields $\tilde{\alpha}= 1$ for all $p < \alpha \leq 1$, which
means that $P_1(p,p_m)=0$ at the beginning of each update.
Therefore, we focus on the case when the gap appears at sites $2
\le i \le L$. Then, the right edge of the gap, positioned at site
$i+1< L$ can move one site to the right, provided that site is
empty. The latter event occurs only if the particle at site $i=L$
leaves the system with probability $\beta$, and the remaining
cluster of $L-i-1$ particles at sites $i+1, \dots, L-1$ moves as a
whole entity one site to the right, which happens with probability
$p_m^{L-i-1}$. Thus, the total probability for the particle at the
right edge to hop to the right is $p_m^{L-i-1}\beta$, and to
remain at its place is $(1-\beta) + (1-p_m^{L-i-1})\beta$. On the
other hand, the particle at the left edge $i-1$, being either the
rightmost particle of a cluster or isolated,  may hop to the right
with the position independent probability $p$, and stay immobile
with probability $1-p$. As a result, the gap width increases by
one site with probability \beq p_g(i)= (1-p)p_m^{L-i-1}\beta,
\label{pg} \eeq decreases by one site with probability \beq
q_g(i)= [(1-\beta) + (1-p_m^{L-i-1})\beta]p =
(1-p_m^{L-i-1}\beta)p, \label{qg} \eeq and remains the same with
probability \beq r_g(i)= 1-p  +p_m^{L-i-1}\beta(2p -1). \label{rg}
\eeq As expected, at $p_m =1$ these expressions reduce to
equalities (4) in Ref. \cite{BBPP18}. As is readily seen, in the
alternative case of several coexisting gaps, the above
probabilities apply exactly to the rightmost one.

Now we have to average the gap width evolution over the initial
probabilities given by (\ref{probgap}). The probability
normalization factor under the condition of a single gap opened at
sites $i=2,3,\dots, L$ is \beq N(p,p_m) = (1- p_m)\beta \sum_{k =
0}^{L-2}p_m^k  = \beta(1- p_m^{L-1}). \eeq

Then, the changes in the gap width at the first time step,
averaged  over all events of gap appearance at sites $i=2,3,\dots,
L$, become as follows:

The gap width increases by one site with probability \beq
\bar{p}_g = \frac{(1-p)(1- p_m)\beta}{p_m(1-p_m^{L-1})}
\sum_{k=2}^{L}p_m^{2k} =
\frac{(1-p)(1+p_m^{L-1})\beta}{p_m(1+p_m)}, \label{pgav} \eeq
decreases by one site with probability \beq \bar{q}_g= \frac{p(1-
p_m)}{(1- p_m^{L-1})}\sum_{k=0}^{L-2}(p_m^k-p_m^{2k-1}\beta) = p -
\frac{p(1+ p_m^{L-1})\beta}{p_m(1+p_m)}, \label{qgav} \eeq and
remains the same with probability \beq \bar{r}_g= \frac{(1-
p_m)}{(1- p_m^{L-1})}\sum_{k=0}^{L-2}[(1-p)p_m^k
+p_m^{2k-1}\beta(2p -1)]= 1-p +
\frac{(2p-1)\beta(1+p_m^{L-1})}{p_m(1+p_m)}. \label{rgav} \eeq
Notably, at $p_m = 1$  the above results reduce again to
equalities (4) in Ref. \cite{BBPP18}.

By comparing the expressions for $\bar{p}_g$ and $\bar{q}_g$, we
conclude that on the average a  single-site gap will grow after
the first time step of its evolution when \beq \beta >
p\frac{p_m(1+p_m)}{(1+p_m^{L-1})}. \eeq When $p_m \rightarrow 1$
and $L$ is fixed, or $L \rightarrow \infty$ so that $p_m^L
\rightarrow 1$, this condition simplifies to $\beta > p$. However,
for fixed values of $p_m$ close to 1, $p_m^L$ will decrease to
zero as $L \rightarrow \infty$. For example, in our computer
simulations we used: $p_m= 0.99$ and $L=800$, which yields
$p_m^{L-1} \simeq 3.22\times 10^{-4}$ and the criterion becomes
much stronger, $\beta > 1.97 p$. However, with each time step $j =
1,2,\dots$ the right edge of the gap will hop forward by one site
with increasing probability $p_m^{L-i-j}\beta$, while its left
edge may hop to the right with the position independent
probability $p$. Thus, the value of $p_g(i)$ will increase and the
value of $q_g(i)$ will decrease in the course of time.

Without going into the involved details of the complete gaps
evolution,  we conjecture that the simple criteria $\beta > p$ for
growing gaps, and $\beta <p$ for decreasing gaps, hold true. Thus,
our expectation, confirmed by the computer simulations, is that in
the upper region  $(p<\alpha \leq 1]\times (p <\beta \leq 1]$ of
the CF phase a maximum-current phase will appear. Its local
density profile satisfies the inequalities $\rho_1 = 1> \rho_{l/2}
> \rho_L$, which follow from the conditions $\tilde{\alpha} =1$,
and the larger probability of gap formation near the end of the
chain. In the lower region $(p < \alpha \leq 1]\times (0 <\beta <
p]$ of the CF phase
 the gaps are scarce, small and short-living, which is indicative
of a high-density phase. Again, the left-hand side of the local
density profile bends upward to $\rho_1 =1$.

Note that in the above consideration  $p = \lim_{p_m \rightarrow
1-0}\sigma_c(p,p_m)$. In the case of $p_m <1$, the critical values
should decrease down to $\sigma_c(p,p)=1-\sqrt{1-p}$, as $p_m
\rightarrow p+0$.

\subsection{Phase diagram and phase transitions}

We performed Monte Carlo simulations of the gTASEP  on open chains
of mainly $L= 800$ and $L= 1600$ sites. Each run started with
$10^6$ relaxation updates and had not less than $10^4$ attempted
updates per lattice site. The stationary properties were evaluated
by averaging over 100 (quasi)independent runs. The estimated
accuracy is $O(10^{-3})$ for the local particle density and
$O(10^{-4})$ for the current.

\begin{figure}[b]
\includegraphics[width=0.6\textwidth,clip]{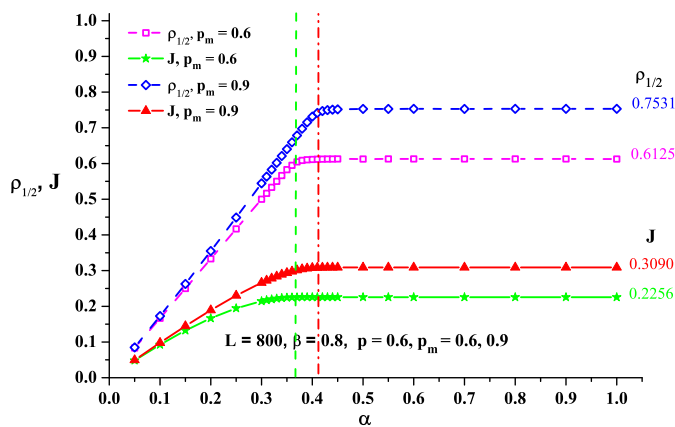}
  \caption{(Color online) Behavior of the stationary midpoint
 density, $\rho_{1/2}$, and the current, $J$, for
 the gTASEP for two values of the modified hopping probability,
  $p_m = 0.6$ and $p_m = 0.9$, as a
 function of the input rate $\alpha$ at chain length $L = 800$ sites,
 fixed $p=0.6$ and output rate $\beta=0.8$. Apparently, their
 behavior (for $p_m = 0.6$ and for $p_m = 0.9$)
  is similar and reflects the continuous nonequilibrium phase
  transition across the segment $\beta_c(p,p_m) < \beta \leq 1$,
  however, with different, $p_m$-dependent critical values: our estimates
  for $p_m = 0.9$ are $\alpha_c(0.6,0.9) = \beta_c(0.6,0.9) \simeq
  0.41$ (shown by the vertical red dashed-dotted line).} \label{Fig2}
\end{figure}

First, we compare the behavior of the current, $J$, and the local
density at the midpoint of the chain, $\rho_{1/2}$, under two
modified hopping probabilities $p_m = 0.6$ and $p_m = 0.9$, as a
function of the input rate $\alpha$, at chain length $L = 800$
sites, fixed $p=0.6$ and output rate $\beta =0.8$, see
Fig.~\ref{Fig2}.

We recall that in the standard backward-sequential TASEP with
$p=0.6$, the exact results in the thermodynamic limit
$L\rightarrow \infty$ are: for the critical injection/ejection
values
 $$\alpha_c =\beta_c = 1 - \sqrt{1-p} = 0.367544\dots, $$
 for the current in the maximum-current phase
 $$J^{MC} = \frac{1-\sqrt{1-p}}{1+\sqrt{1-p}}= 0.225148\dots ,$$
 and for the midpoint density in the MC phase
 $$\rho_{1/2}^{MC} = \frac{1}{1+\sqrt{1-p}}= 0.612574\dots .$$
 The critical value $\alpha_c$, shown in Fig. \ref{Fig2} by a vertical
 green dashed line, corresponds to the transition of the asymptotic
 behavior of the current $J$ (at $p_m=0.6$) near $\alpha_c$ from a parabolic one on
 the left-hand side of the segment $\beta_c(0.6,0.6) < \beta \leq 1$,
 to a constant value in the MC phase on the right-hand side of it.
 Due to finite-size effects, the current $J^{MC}$ slightly grows
 with $\alpha$ up to $0.2256$ at $\alpha =1$.

Similarly, at $p_m = 0.9$, we see that the phase transition across
the segment $\beta_c(0.6,0.9) < \beta \leq 1$ is continuous too
and we estimate the critical values $\alpha_c(0.6,0.9) =
\beta_c(0.6,0.9)\simeq 0.41$, see the vertical red dash-dotted
line in Fig. \ref{Fig2}. This indicates that the unusual phase
transition, found in \cite{BunP17} at $p_m = 1$ across the
boundary $\alpha = p$ becomes a continuous one. Note that in the
MC phase the current grows up to $J^{MC}(0.6,0.9) = 0.3090\dots$
and the local density $\rho^{MC}_{1/2}(0.6,0.9)$ up to 0.7531.

\begin{figure}[t]
\includegraphics [width=.475\textwidth]{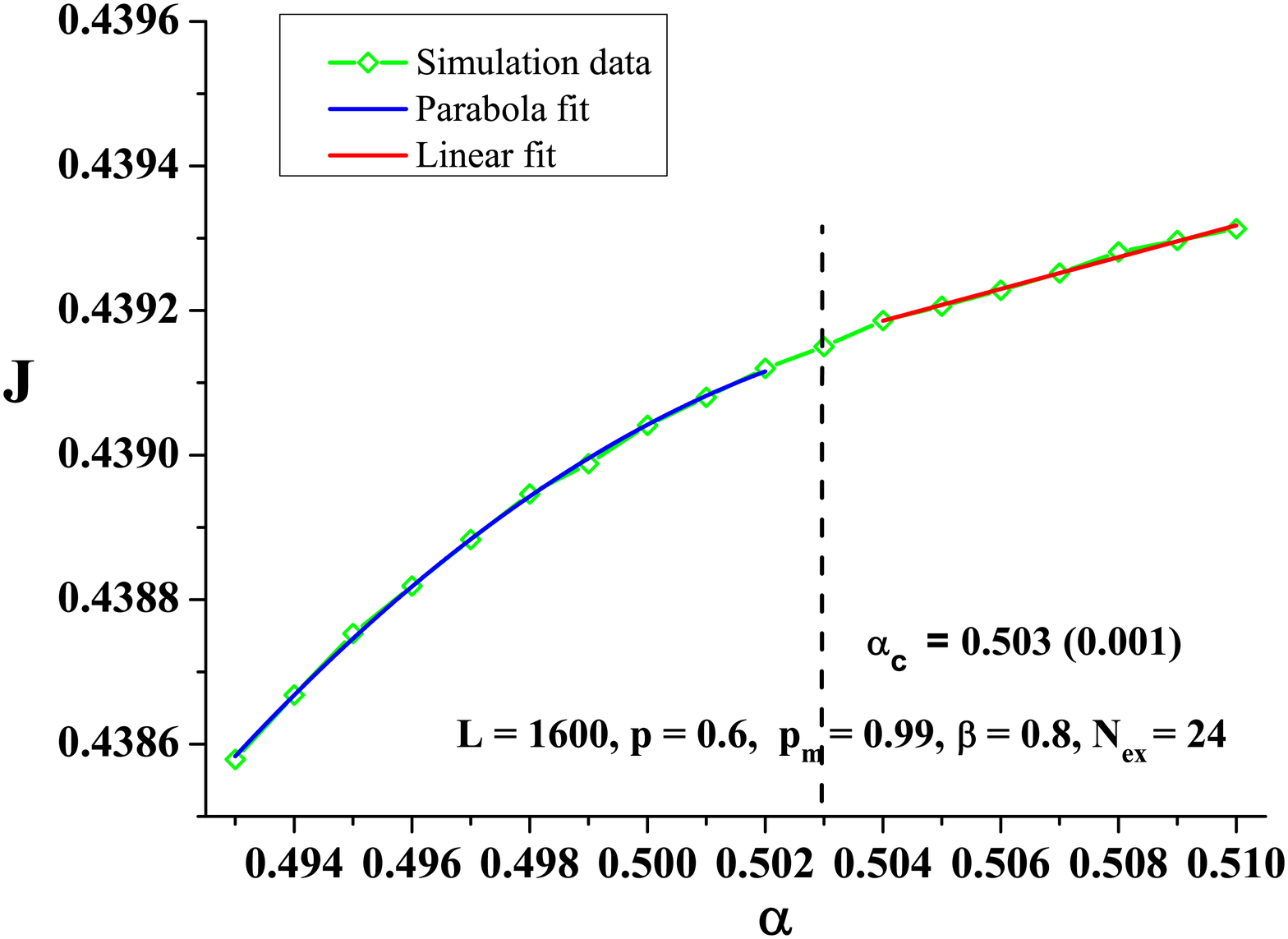}  \quad  
\includegraphics [width=.475\textwidth]{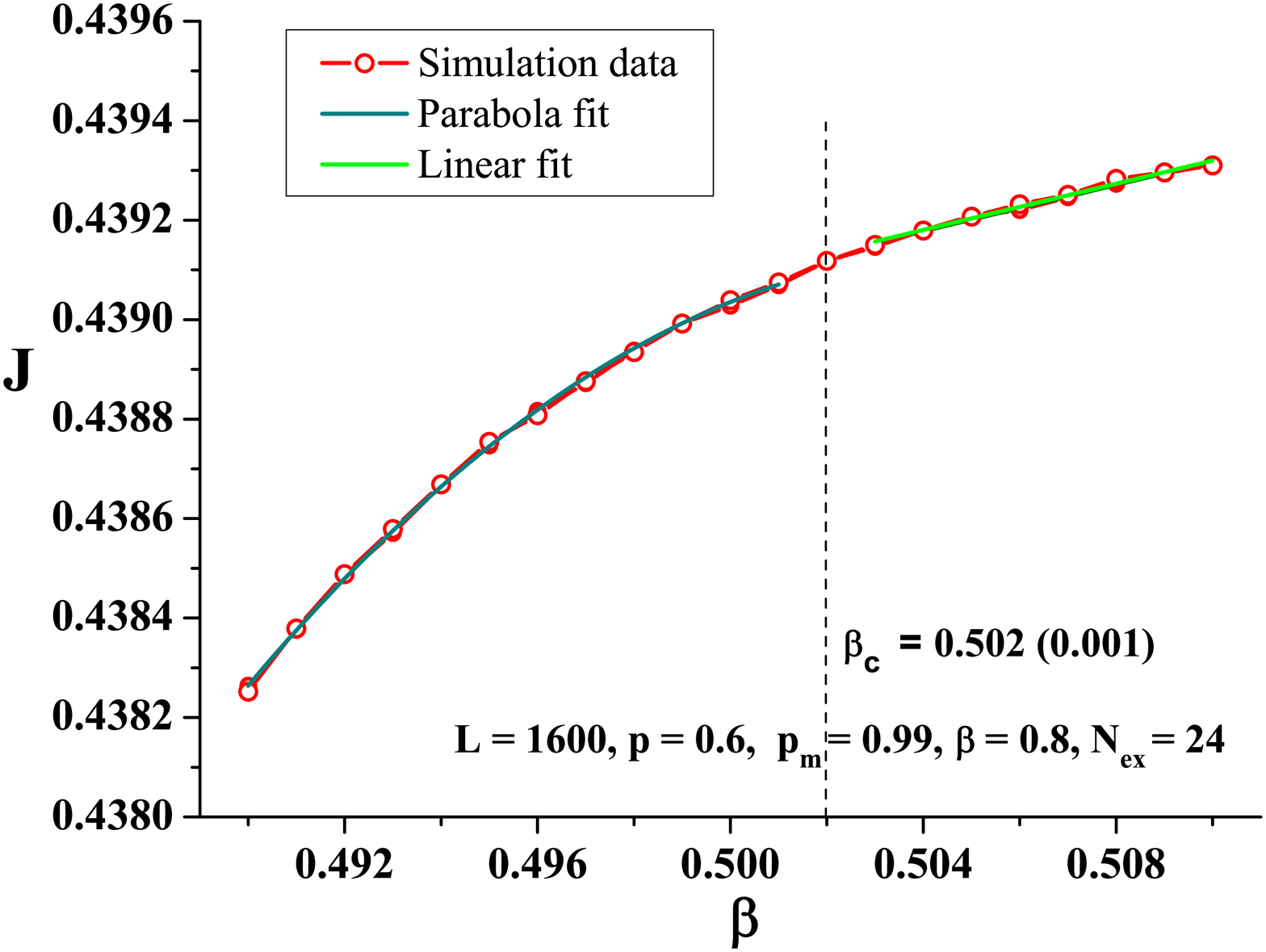}  \\ 
\quad (a) \hspace{.4\textwidth} (b) \caption{(Color online) The
current in the gTASEP when $p=0.6$, $p_m=0.99$ and $L=1600$ sites,
as a function of: (a) the injection probability $\alpha$ (at
$\beta = 0.8$). The critical value $\alpha_c(0.6,0.99)= 0.503\pm
0.001$ is estimated from the apparent change in the asymptotic
behavior of the current; (b) the ejection probability $\beta$ (at
$\alpha =0.8$). The critical value $\beta_c(0.6,0.99)= 0.502\pm
0.001$ is estimated from the apparent change in the asymptotic
behavior of the current.} \label{J99AB08}
\end{figure}

To check the continuity of the phase transition across the segment
$\beta_c < \beta \leq 1$, we consider in more detail both the
$\alpha$- and $\beta$-dependance of the current on larger lattice
and value of $p_m$ closer to 1, namely $L=1600$ and $p_m  =0.99$.
The results are shown in Fig. \ref{J99AB08}(a) and (b).

It is instructive to analytically check the above estimates for
the critical values of the injection (ejection) $\alpha_c$
($\beta_c$) probabilities, commonly denoted by $\sigma_c$. To this
end we use the continuity condition for the first derivative of
the current: $J'(\sigma_c -0)=J'(\sigma_c +0)$. Close to the
critical value $\sigma_c$, we have a parabolic approximation for
the current below $\sigma_c$, \beq J(\sigma <\sigma_c)= A +
B\sigma + C\sigma^2, \label{parab} \eeq and a linear approximation
above $\sigma_c$, \beq J(\sigma >\sigma_c)= a + k\sigma.
\label{lin} \eeq Hence, in the case of a continuous second order
phase transition,  when $J'(\sigma_c -0)=J'(\sigma_c +0)$, we
obtain an estimate for $\sigma_c$: \beq \sigma_c = \frac{k-B}{2C}.
\eeq In the case of the current as a function of $\alpha$, see
Fig. \ref{J99AB08}(a), the best least-square fit yields: \beq A =
- 0.386 \pm 0.05,\quad B = 3.229 \pm 0.21,\quad C=-3.186 \pm 0.21,
\quad k = 0.022 \pm 7.7 \times 10^{-4}, \eeq which leads to the
estimate $\alpha_c = 0.503 \pm 0.07$. In spite of the large error
interval, this estimate coincides with our former value of
$\alpha_c(0.6,0.99)\simeq 0.503$.

In the complementary case of the $\beta$-dependent current,  see
Fig. \ref{J99AB08}(b), the best least-square fit yields: \beq A =
- 0.536 \pm 0.055,\quad B = 3.86 \pm 0.22,\quad C=-3.82 \pm 0.22,
\quad k = 0.023 \pm 0.001, \eeq which leads to the estimate
$\beta_c = 0.502 \pm 0.05$. Again, the error bars are rather
large, but this estimated value also coincides with our former
assessment of $\beta_c(0.6,0.99)\simeq 0.502$. Finally, we assume
that within error bars $\alpha_c(0.6,0.99)=\beta_c(0.6,0.99)=
0.502\pm 0.02$.

\begin{figure}[t]
\includegraphics[width=0.6\textwidth,clip]{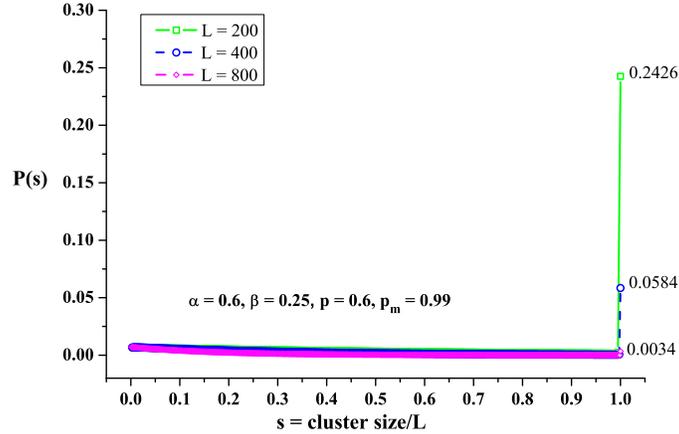} 
\caption{(Color online) Cluster size distribution in the
 gTASEP when $p = 0.6$, $p_m = 0.99$ on lattices $L=200, 400, 800$ sites
 at $\alpha =0.6$, and $\beta = 0.25$.
At $L = 1600$ the estimated  value of $P(1)$ drops down to $1.147
\times 10^{-5}$.} \label{ClDist025}
\end{figure}

Now we focus on the phase transitions taking place by changing
$\alpha$  across the segment $0< \beta < \beta_c(p,p_m)$ . In Ref.
\cite{BunP17}, a mixed MP+CF phase was found at $p_m = 1$, see
Fig. \ref{PhD}(b). This phase is characterized by the nonzero
probability $P(1)$ of appearance of a cluster spanning the whole
chain of $L$ sites: $P(1)$ changes with $\alpha$ from zero at the
left phase boundary $0 < \alpha =\beta < p$ to $P(1)=1$ at the
right boundary $\alpha = p$, $0 <\beta < p$ with the CF phase. In
Ref. \cite{BBPP18} the MP+CF phase was interpreted as a boundary
perturbed one. Here we show that as $p_m <1$, $P(1)$ exponentially
decreases to zero, not only in the MP+CF phase but also in the
subregion $(0< \alpha < \alpha_c)\times (0< \beta < \beta_c)$,
which at $p_m = 1$ belongs to the CF phase. The results of our
computer simulations for the cluster-size distribution in gTASEP
when $p_m = 0.99$ at $\alpha = 0.6$, $\beta = 0.25$,  and lattice
sizes $L=200, 400, 800$ are shown in Fig. \ref{ClDist025}.

\begin{figure}[b]
\includegraphics[width=0.6\textwidth,clip]{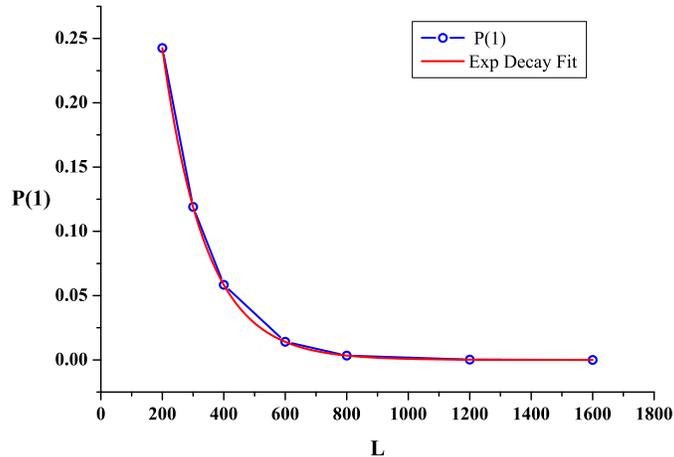} 
\caption{(Color online)  Least-square fit to the exponential decay
of the complete-cluster probability $P(1)$ with the unlimited
increase of the chain length $L$ in the gTASEP when $p = 0.6$ and
$p_m = 0.99$ at $\alpha =0.6$ and $\beta = 0.25$.}
\label{ExpDecay}
\end{figure}
By using a larger series of chain sizes, from $L=200$ to $L=1600$,
we obtain that $P(1)$ probability decays exponentially fast with
the unlimited increase of $L$, \beq P_L(1) \simeq 0.2426\times
\exp\{-(L-200)/140\}, \quad L\geq 200. \eeq The quality of the fit
is illustrated in Fig. \ref{ExpDecay}.

\begin{figure}[t]
\includegraphics[width=0.6\textwidth,clip]{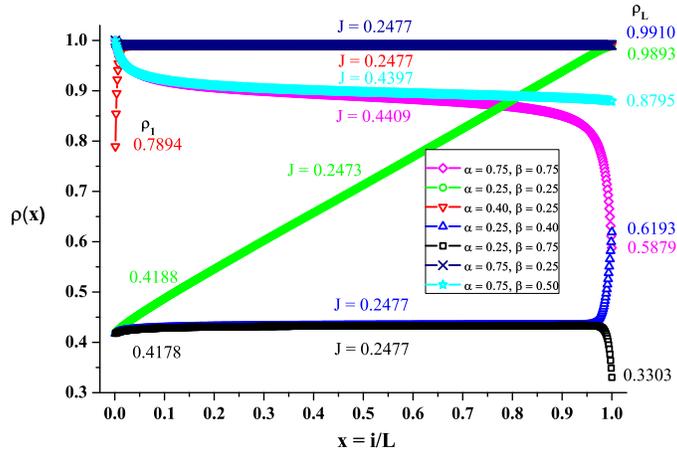}
\caption{(Color online) Local density profiles of the gTASEP when
$p = 0.6$ and $p_m = 0.99$ at different points in the phase space.
Obviously, the point $(\alpha =0.25,\beta = 0.25)$ lies on the
coexistence line between the low-density phase, represented by the
points $(\alpha =0.25,\beta = 0.40)$ and $(\alpha =0.25,\beta =
0.75)$, and the high-density phase, represented by the points
$(\alpha =0.40,\beta = 0.25)$, $(\alpha =0.75,\beta = 0.25)$ and
$(\alpha =0.75,\beta = 0.50)$. The corresponding value of the
current $J$ is denoted next to every density profile.}
\label{Ro99DifPhases}
\end{figure}

Therefore, by continuity arguments, we conclude that in the
thermodynamic  limit $L \rightarrow \infty$ the regions between
the left-hand  boundary $0 < \alpha =\beta < \sigma_c(p, p_m)$ and
the right-hand boundary at $\alpha =1$ and $0 < \beta < \beta_c(p,
p_m)$, belong to the same phase. The fact that across the
coexistence line $0 < \alpha =\beta < \sigma_c(p, p_m)$ there
occurs a first-order phase transition is seen from the shape of
the local density profiles, shown in Fig.~\ref{Ro99DifPhases} at
different points $(\alpha, \beta)$ in the $\alpha - \beta$ plane.

\begin{figure}[t]
\includegraphics [width=0.5\textwidth]{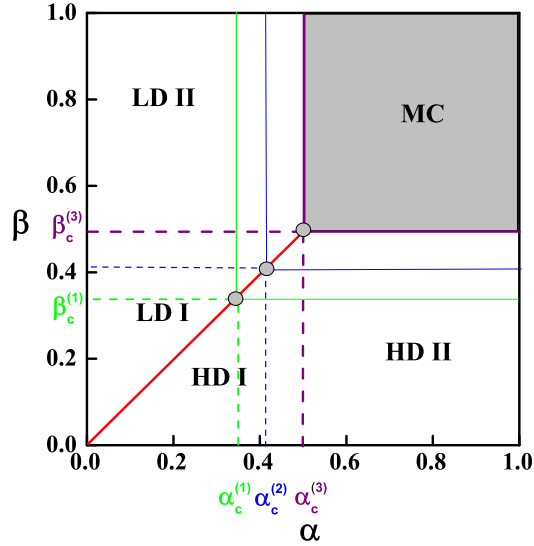}
\caption{\label{PhD99} (Color online) Conjectured phase diagram of
the gTASEP when $p = 0.6$ and $p_m = 0.99$ (purple lines). It has
the same topology as in the case of the standard TASEP with
backward-sequential update (shown in Fig. 1(a)), but with
different, $p_m$-dependent critical values -- our estimates are
$\alpha_c^{(3)}(0.6; 0.99) = \beta_c^{(3)}(0.6; 0.99)=0.502$, see
text. To illustrate the shift of the triple point we have added
also the critical lines corresponding of the TASEP with backward
sequential update, i.e., $\alpha_c^{(1)}(0.6; 0.6) =
\beta_c^{(1)}(0.6; 0.6)=0.3675$ (thin green lines),  and also of
the gTASEP when $p = 0.6$ and $p_m = 0.9$ --
$\alpha_c^{(2)}(0.6;0.9) = \beta_c^{2}(0.6;0.9)=0.41$ (thin blue
lines).}
\end{figure}

By using continuity arguments, we generalize the above results  to
conjecture a generic phase diagram of the gTASEP with $p_m < 1$
with the same topology as in the case of the backward-sequential
TASEP, see Fig. \ref{PhD}(a), but with $(p, p_m)$-dependent triple
point $(\alpha_c,\beta_c)$. In Fig. \ref{PhD99} we exemplify the
phase diagram of the gTASEP in the particular case of $p = 0.6$
and $p_m = 0.99$ and the shift of the triple point
$(\alpha_c,\beta_c)$ with the increase of $p_m$ at fixed $p=0.6$.

Now one can see the similarity in the behavior of the local
density profiles  in the cases $p_m = p <1$ and $p < p_m < 1$. In
the low-density phase LD = LDI$\cup$LDII the bulk density is less
than 1/2, the difference between the LDI and LDII regions is in
the right-hand end of the local density profile: in LDI it bends
upward, while in LDII it bends downward, similarly to the case of
the standard backward-sequential TASEP. This can be readily
explained by using the exact relationship $\rho_L =J/\beta$, and
assuming the same value of the current in the two regions.  In the
considered particular case of $p=0.6$ and $p_m =0.99$, the bulk
density in the high-density phase is very close to 1, the
difference between the regions HDI and HDII being at left-hand
side of the profile: in HDI it sharply bends downward, while in
HDII it bends upward, again similarly to the case of the standard
backward-sequential TASEP.

\begin{figure}[t]
\includegraphics [width=.45\textwidth]{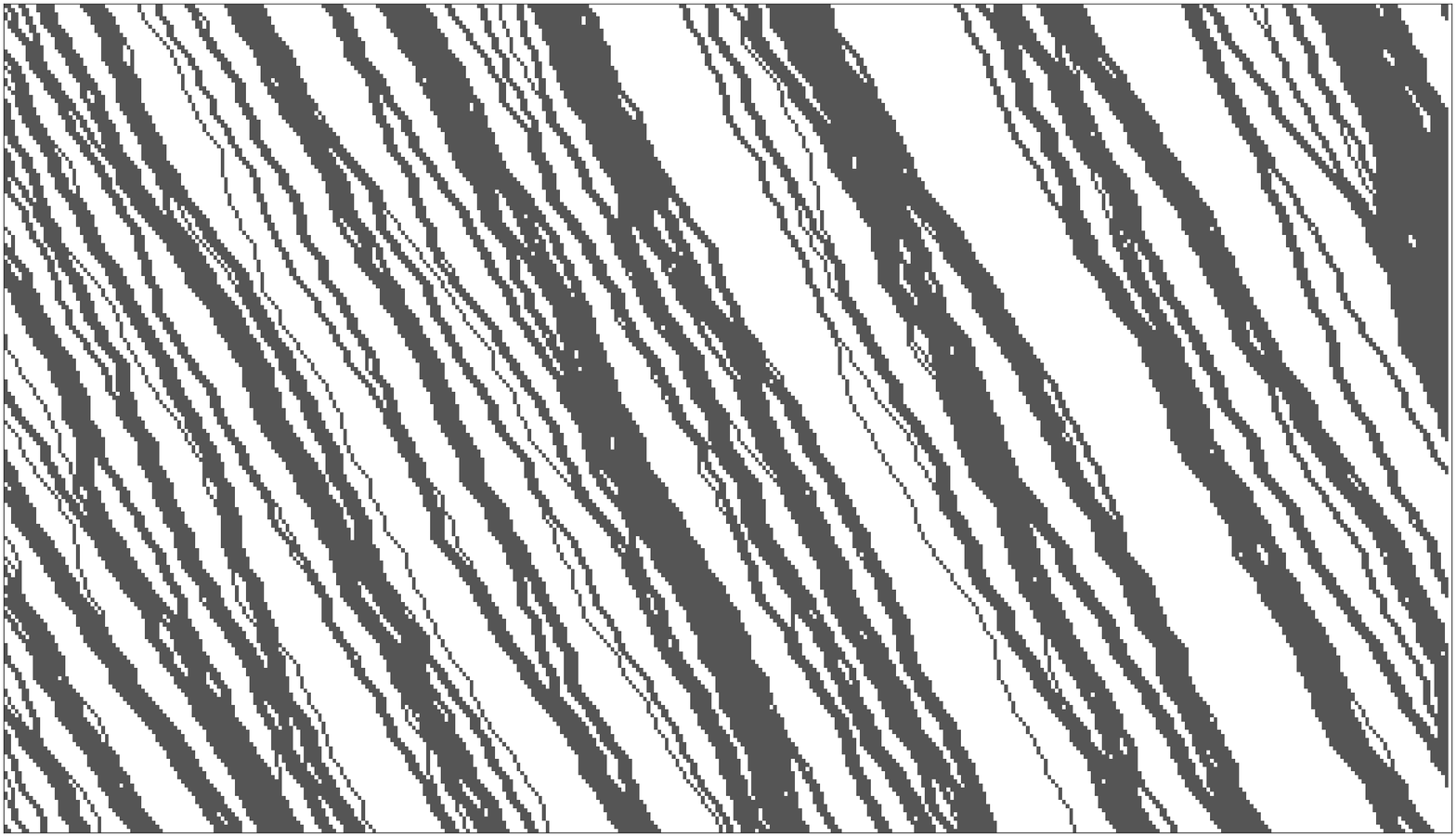}  \quad  
\includegraphics [width=.45\textwidth]{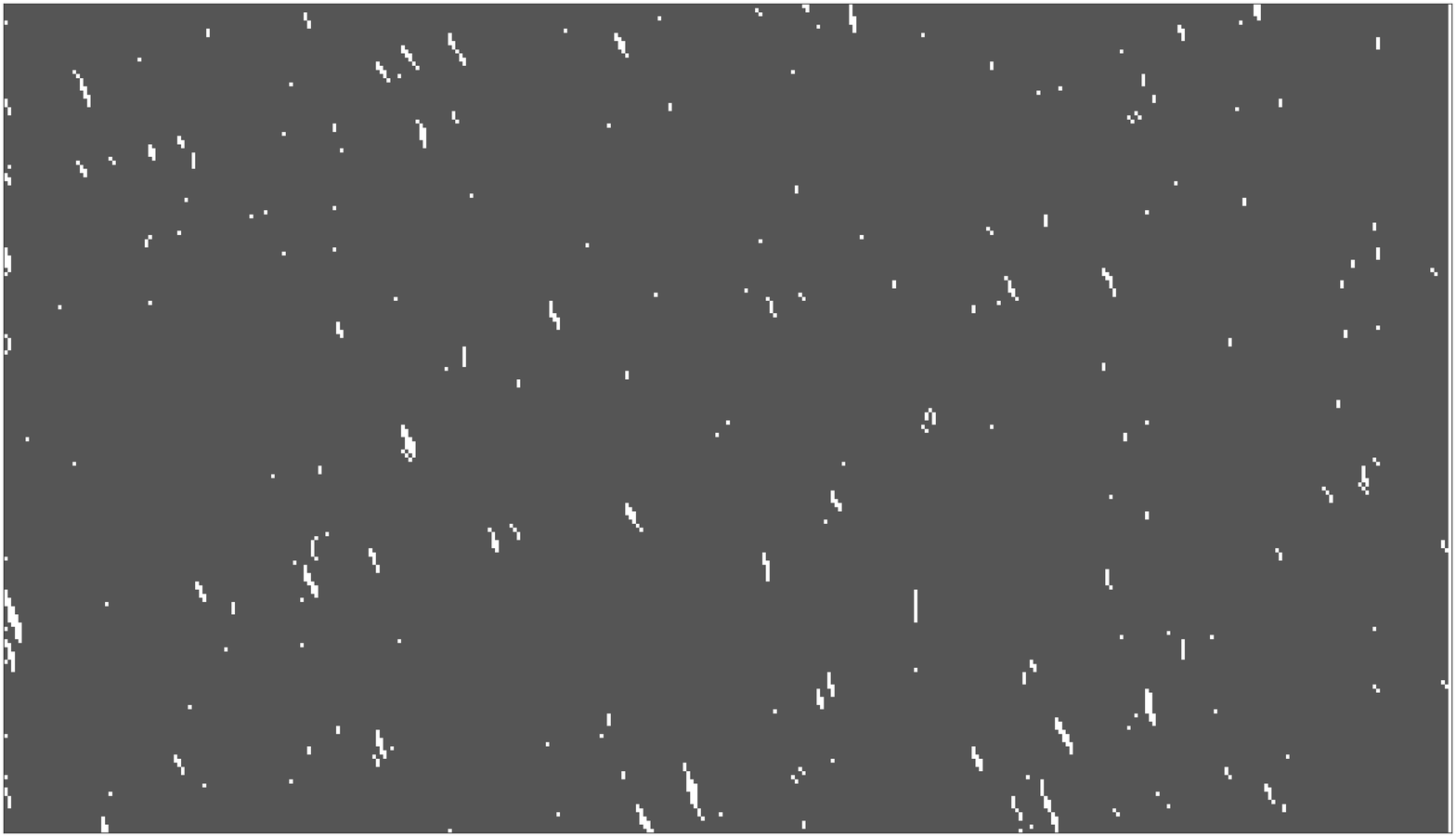}  \\ 
\quad (a) \hspace{.45\textwidth} (b) \caption{ A space-time plot
(time is flowing downward in the vertical direction) of the gTASEP
when $p=0.6$, $p_m=0.99$, and $L=400$ sites, showing the gaps
evolution,  in: (a) region LDI ($\alpha = 0.25$,$\beta = 0.40$) --
white stripes; (b) in region HDI ($\alpha = 0.40$, $\beta = 0.25$)
-- white spots.} \label{Gaps99a2540}
\end{figure}
Additional information can be found in the different gaps
evolution regimes in regions LDI and HDI: in both cases $\alpha
<p$, which implies $\tilde{\alpha} < 1$, so that gaps can appear
at the first site $i=1$ and evolve throughout the chain; however,
in LDI the gaps are wide and long-living, while in HDI they are
small, scarce and very short living, compare Figs.
\ref{Gaps99a2540}(a) and \ref{Gaps99a2540}(b). The typical gaps
pattern in LDII (HDII)
is similar to the one shown for LDI (HDI). These features may   
explain the large difference in the particle densities in the
low-density and high-density phases.

Here we emphasize that the  gTASEP does not satisfy the
particle-hole symmetry inherent to the standard versions of TASEP.
For example, the fundamental diagram of the generic model,
obtained by Hrab{\'a}k, see Fig. 6.2 in \cite{PhD}, where $\gamma
= p_m/p$, is not symmetric under the replacement $\rho
\leftrightarrow 1- \rho$. The above mentioned phase diagram was
obtained in the thermodynamic limit under periodic boundary
conditions, which means that the very bulk dynamics at $p <p_m$
does not respect the particle-hole symmetry. In addition, in our
case of open boundaries, the left boundary condition Eq.
(\ref{lbc}) is not appropriate for introduction of holes from the
right chain end. Therefore, it is rather unexpected that the
currents in the phases LD and HD, calculated at symmetric points
$(\alpha,\beta)$ and $(\beta, \alpha)$,  are equal: $J^{LD}(\alpha
=0.25, \beta =0.75) = J^{HD}(\alpha =0.75, \beta =0.25) \simeq
0.2477$.

\section{Discussion}

We studied the generalized TASEP in the regime of particle
attraction  ($p_m >p$) between hopping nearest-neighboring
particles. In this case ($p<p_m<1$) cluster aggregation and
fragmentation is allowed in the system. A central problem of
interest was to find how the topology of the phase diagram in the
case of irreversible aggregation $p_m =1$, see Fig. \ref{PhD}(b),
transforms into the topology of the well-known phase diagram of
the usual TASEP with backward-ordered sequential update, see Fig.
\ref{PhD}(a), when $p_m$ decreases from $p_m =1$ down to $p_m =p$.
Based on an incomplete random-walk theory and on extensive Monte
Carlo calculations, we conjectured that the above phenomenon takes
place sharply, as soon as $p_m$ becomes less than 1. The main
difference between the phase diagrams for $p_m =1$ and $p_m <1$
turned out to be the dependence of the critical probabilities
$\sigma_c(p,p_m)$ on $p_m$, at fixed $p$. Apart of that, we have
shown the similarity of the local density profiles and the current
as a function of the injection $\alpha$ and ejection $\beta$
probabilities, in the cases $p_m =p$ and $p_m
>p$. The main effect of increasing the modified hopping
probability $p_m$  turns out to be increase in the values of
critical point coordinates, the bulk density and the current. For
example, on passing from $p_m=p$ to $p_m=0.99$, these values grow
from $$\alpha_c =\beta_c \simeq 0.3675, \quad J^{\text{MC}} \simeq
0.2251, \quad \rho_{1/2}^{\text{MC}} \simeq 0.6126,$$ to
$$\alpha_c =\beta_c \simeq 0.502, \quad J^{\text{MC}} \simeq 0.4409,
\quad \rho_{1/2}^{\text{MC}} \simeq 0.8891.$$ On the ground of our
random walk  theory and the computer simulations, we have
conjectured that the simple criteria $\beta >p$, for growing gaps,
and $\beta < p$, for decreasing gaps, hold true on the average.

An interesting result is the exponential decay to zero of the
probability $P(1)$ of a complete cluster in the HDII phase (which
emerges in the lower region of the CF phase of the gTASEP with
$p_m=1$), when $p_m< 1$ and the chain length $L$ increases
unboundedly, see Fig.~\ref{ExpDecay}.

There are still many open problems, such as an elaboration of the
random walk  theory to the extend of yielding both qualitative and
quantitative predictions, the analytical derivation of the local
density at the chain ends and the value of the current in the
different stationary phases, just to mention some.

\begin{acknowledgments}
The authors gratefully acknowledge the fruitful discussions with
their  late colleague and coauthor Professor V.B. Priezzhev, held
at an early stage 
of the present study. NCP also acknowledges the provided access to
the e-infrastructure of the Centre for Advanced Computing and Data
Processing, with the financial support by the Grant No
BG05M2OP001-1.001-0003, financed by the Science and Education for
Smart Growth Operational Program (2014-2020) and co-financed by
the European Union through the European structural and Investment
funds.

\end{acknowledgments}

\bibliography{AggregFragmBibl}

\begin{thebibliography}{27}%
\makeatletter
\providecommand \@ifxundefined [1]{%
 \@ifx{#1\undefined}
}%
\providecommand \@ifnum [1]{%
 \ifnum #1\expandafter \@firstoftwo
 \else \expandafter \@secondoftwo
 \fi
}%
\providecommand \@ifx [1]{%
 \ifx #1\expandafter \@firstoftwo
 \else \expandafter \@secondoftwo
 \fi
}%
\providecommand \natexlab [1]{#1}%
\providecommand \enquote  [1]{``#1''}%
\providecommand \bibnamefont  [1]{#1}%
\providecommand \bibfnamefont [1]{#1}%
\providecommand \citenamefont [1]{#1}%
\providecommand \href@noop [0]{\@secondoftwo}%
\providecommand \href [0]{\begingroup \@sanitize@url \@href}%
\providecommand \@href[1]{\@@startlink{#1}\@@href}%
\providecommand \@@href[1]{\endgroup#1\@@endlink}%
\providecommand \@sanitize@url [0]{\catcode `\\12\catcode `\$12\catcode
  `\&12\catcode `\#12\catcode `\^12\catcode `\_12\catcode `\%12\relax}%
\providecommand \@@startlink[1]{}%
\providecommand \@@endlink[0]{}%
\providecommand \url  [0]{\begingroup\@sanitize@url \@url }%
\providecommand \@url [1]{\endgroup\@href {#1}{\urlprefix }}%
\providecommand \urlprefix  [0]{URL }%
\providecommand \Eprint [0]{\href }%
\providecommand \doibase [0]{http://dx.doi.org/}%
\providecommand \selectlanguage [0]{\@gobble}%
\providecommand \bibinfo  [0]{\@secondoftwo}%
\providecommand \bibfield  [0]{\@secondoftwo}%
\providecommand \translation [1]{[#1]}%
\providecommand \BibitemOpen [0]{}%
\providecommand \bibitemStop [0]{}%
\providecommand \bibitemNoStop [0]{.\EOS\space}%
\providecommand \EOS [0]{\spacefactor3000\relax}%
\providecommand \BibitemShut  [1]{\csname bibitem#1\endcsname}%
\let\auto@bib@innerbib\@empty
\bibitem [{\citenamefont {MacDonald}\ \emph {et~al.}(1968)\citenamefont
  {MacDonald}, \citenamefont {Gibbs},\ and\ \citenamefont {Pipkin}}]{MGP68}%
  \BibitemOpen
  \bibfield  {author} {\bibinfo {author} {\bibfnamefont {C.~T.}\ \bibnamefont
  {MacDonald}}, \bibinfo {author} {\bibfnamefont {J.~H.}\ \bibnamefont
  {Gibbs}}, \ and\ \bibinfo {author} {\bibfnamefont {A.~C.}\ \bibnamefont
  {Pipkin}},\ }\href@noop {} {\bibfield  {journal} {\bibinfo  {journal}
  {Biopolymers}\ }\textbf {\bibinfo {volume} {6}},\ \bibinfo {pages} {1}
  (\bibinfo {year} {1968})}\BibitemShut {NoStop}%
\bibitem [{\citenamefont {Zhang}\ \emph {et~al.}(2006)\citenamefont {Zhang},
  \citenamefont {Liu},\ and\ \citenamefont {Hu}}]{ZLH}%
  \BibitemOpen
  \bibfield  {author} {\bibinfo {author} {\bibfnamefont {F.}~\bibnamefont
  {Zhang}}, \bibinfo {author} {\bibfnamefont {C.~L.}\ \bibnamefont {Liu}}, \
  and\ \bibinfo {author} {\bibfnamefont {B.~R.}\ \bibnamefont {Hu}},\
  }\href@noop {} {\bibfield  {journal} {\bibinfo  {journal} {J.
  Neurochemistry}\ }\textbf {\bibinfo {volume} {4}},\ \bibinfo {pages} {102}
  (\bibinfo {year} {2006})}\BibitemShut {NoStop}%
\bibitem [{\citenamefont {Tipathi}\ and\ \citenamefont
  {Chowdhury}(2008)}]{TC08}%
  \BibitemOpen
  \bibfield  {author} {\bibinfo {author} {\bibfnamefont {T.}~\bibnamefont
  {Tipathi}}\ and\ \bibinfo {author} {\bibfnamefont {D.}~\bibnamefont
  {Chowdhury}},\ }\href@noop {} {\bibfield  {journal} {\bibinfo  {journal}
  {Phys. Rev. E}\ }\textbf {\bibinfo {volume} {77}},\ \bibinfo {pages} {011921}
  (\bibinfo {year} {2008})}\BibitemShut {NoStop}%
\bibitem [{\citenamefont {Guelich}\ \emph {et~al.}(2007)\citenamefont
  {Guelich}, \citenamefont {Garai}, \citenamefont {Nashinari}, \citenamefont
  {Schadshneider},\ and\ \citenamefont {Chowdhury}}]{GGNSC07}%
  \BibitemOpen
  \bibfield  {author} {\bibinfo {author} {\bibfnamefont {P.}~\bibnamefont
  {Guelich}}, \bibinfo {author} {\bibfnamefont {A.}~\bibnamefont {Garai}},
  \bibinfo {author} {\bibfnamefont {K.}~\bibnamefont {Nashinari}}, \bibinfo
  {author} {\bibfnamefont {A.}~\bibnamefont {Schadshneider}}, \ and\ \bibinfo
  {author} {\bibfnamefont {D.}~\bibnamefont {Chowdhury}},\ }\href@noop {}
  {\bibfield  {journal} {\bibinfo  {journal} {Phys. Rev. E}\ }\textbf {\bibinfo
  {volume} {75}},\ \bibinfo {pages} {041905} (\bibinfo {year}
  {2007})}\BibitemShut {NoStop}%
\bibitem [{\citenamefont {Kolomeisky}(2007)}]{K07}%
  \BibitemOpen
  \bibfield  {author} {\bibinfo {author} {\bibfnamefont {A.~B.}\ \bibnamefont
  {Kolomeisky}},\ }\href@noop {} {\bibfield  {journal} {\bibinfo  {journal}
  {Phys. Rev. Lett.}\ }\textbf {\bibinfo {volume} {98}},\ \bibinfo {pages}
  {048105} (\bibinfo {year} {2007})}\BibitemShut {NoStop}%
\bibitem [{\citenamefont {Zilman}\ \emph {et~al.}(2009)\citenamefont {Zilman},
  \citenamefont {Pearson},\ and\ \citenamefont {Bel}}]{ZPB09}%
  \BibitemOpen
  \bibfield  {author} {\bibinfo {author} {\bibfnamefont {A.}~\bibnamefont
  {Zilman}}, \bibinfo {author} {\bibfnamefont {J.}~\bibnamefont {Pearson}}, \
  and\ \bibinfo {author} {\bibfnamefont {G.}~\bibnamefont {Bel}},\ }\href@noop
  {} {\bibfield  {journal} {\bibinfo  {journal} {Phys. Rev. Lett.}\ }\textbf
  {\bibinfo {volume} {103}},\ \bibinfo {pages} {128103} (\bibinfo {year}
  {2009})}\BibitemShut {NoStop}%
\bibitem [{\citenamefont {Nagle}(1996)}]{N96}%
  \BibitemOpen
  \bibfield  {author} {\bibinfo {author} {\bibfnamefont {K.}~\bibnamefont
  {Nagle}},\ }\href@noop {} {\bibfield  {journal} {\bibinfo  {journal} {Phys.
  Rev. E}\ }\textbf {\bibinfo {volume} {53}},\ \bibinfo {pages} {4655}
  (\bibinfo {year} {1996})}\BibitemShut {NoStop}%
\bibitem [{\citenamefont {Chowdhury}\ \emph {et~al.}(2000)\citenamefont
  {Chowdhury}, \citenamefont {Santen},\ and\ \citenamefont
  {Schadschneider}}]{CSS00}%
  \BibitemOpen
  \bibfield  {author} {\bibinfo {author} {\bibfnamefont {D.}~\bibnamefont
  {Chowdhury}}, \bibinfo {author} {\bibfnamefont {L.}~\bibnamefont {Santen}}, \
  and\ \bibinfo {author} {\bibfnamefont {A.}~\bibnamefont {Schadschneider}},\
  }\href@noop {} {\bibfield  {journal} {\bibinfo  {journal} {Phys. Rep.}\
  }\textbf {\bibinfo {volume} {329}},\ \bibinfo {pages} {199} (\bibinfo {year}
  {2000})}\BibitemShut {NoStop}%
\bibitem [{\citenamefont {Arita}\ \emph {et~al.}(2017)\citenamefont {Arita},
  \citenamefont {Foulaadvand},\ and\ \citenamefont {Santen}}]{AFS17}%
  \BibitemOpen
  \bibfield  {author} {\bibinfo {author} {\bibfnamefont {C.}~\bibnamefont
  {Arita}}, \bibinfo {author} {\bibfnamefont {M.~E.}\ \bibnamefont
  {Foulaadvand}}, \ and\ \bibinfo {author} {\bibfnamefont {L.}~\bibnamefont
  {Santen}},\ }\href@noop {} {\bibfield  {journal} {\bibinfo  {journal} {Phys.
  Rev. E}\ }\textbf {\bibinfo {volume} {95}},\ \bibinfo {pages} {032108}
  (\bibinfo {year} {2017})}\BibitemShut {NoStop}%
\bibitem [{\citenamefont {Helbing}(2001)}]{H01}%
  \BibitemOpen
  \bibfield  {author} {\bibinfo {author} {\bibfnamefont {D.}~\bibnamefont
  {Helbing}},\ }\href@noop {} {\bibfield  {journal} {\bibinfo  {journal} {Rev.
  Mod. Phys.}\ }\textbf {\bibinfo {volume} {73}},\ \bibinfo {pages} {1067}
  (\bibinfo {year} {2001})}\BibitemShut {NoStop}%
\bibitem [{\citenamefont {Derrida}\ \emph {et~al.}(1992)\citenamefont
  {Derrida}, \citenamefont {Domany},\ and\ \citenamefont {Mukamel}}]{DDM92}%
  \BibitemOpen
  \bibfield  {author} {\bibinfo {author} {\bibfnamefont {B.}~\bibnamefont
  {Derrida}}, \bibinfo {author} {\bibfnamefont {E.}~\bibnamefont {Domany}}, \
  and\ \bibinfo {author} {\bibfnamefont {D.}~\bibnamefont {Mukamel}},\
  }\href@noop {} {\bibfield  {journal} {\bibinfo  {journal} {J. Stat. Phys.}\
  }\textbf {\bibinfo {volume} {69}},\ \bibinfo {pages} {667} (\bibinfo {year}
  {1992})}\BibitemShut {NoStop}%
\bibitem [{\citenamefont {Sch\"{u}tz}\ and\ \citenamefont
  {Domany}(1993)}]{SD93}%
  \BibitemOpen
  \bibfield  {author} {\bibinfo {author} {\bibfnamefont {G.~M.}\ \bibnamefont
  {Sch\"{u}tz}}\ and\ \bibinfo {author} {\bibfnamefont {E.}~\bibnamefont
  {Domany}},\ }\href@noop {} {\bibfield  {journal} {\bibinfo  {journal} {J.
  Stat. Phys.}\ }\textbf {\bibinfo {volume} {72}},\ \bibinfo {pages} {277}
  (\bibinfo {year} {1993})}\BibitemShut {NoStop}%
\bibitem [{\citenamefont {Derrida}\ \emph {et~al.}(1993)\citenamefont
  {Derrida}, \citenamefont {Evans}, \citenamefont {Hakim},\ and\ \citenamefont
  {Pasquier}}]{DEHP}%
  \BibitemOpen
  \bibfield  {author} {\bibinfo {author} {\bibfnamefont {B.}~\bibnamefont
  {Derrida}}, \bibinfo {author} {\bibfnamefont {M.~R.}\ \bibnamefont {Evans}},
  \bibinfo {author} {\bibfnamefont {V.}~\bibnamefont {Hakim}}, \ and\ \bibinfo
  {author} {\bibfnamefont {V.}~\bibnamefont {Pasquier}},\ }\href@noop {}
  {\bibfield  {journal} {\bibinfo  {journal} {J. Phys. A}\ }\textbf {\bibinfo
  {volume} {26}},\ \bibinfo {pages} {1493} (\bibinfo {year}
  {1993})}\BibitemShut {NoStop}%
\bibitem [{\citenamefont {Hinrichsen}(1996)}]{H96}%
  \BibitemOpen
  \bibfield  {author} {\bibinfo {author} {\bibfnamefont {H.}~\bibnamefont
  {Hinrichsen}},\ }\href@noop {} {\bibfield  {journal} {\bibinfo  {journal} {J.
  Phys. A}\ }\textbf {\bibinfo {volume} {29}},\ \bibinfo {pages} {3659}
  (\bibinfo {year} {1996})}\BibitemShut {NoStop}%
\bibitem [{\citenamefont {Honecker}\ and\ \citenamefont
  {Peschel}(1997)}]{HP97}%
  \BibitemOpen
  \bibfield  {author} {\bibinfo {author} {\bibfnamefont {A.}~\bibnamefont
  {Honecker}}\ and\ \bibinfo {author} {\bibfnamefont {I.}~\bibnamefont
  {Peschel}},\ }\href@noop {} {\bibfield  {journal} {\bibinfo  {journal} {J.
  Stat. Phys}\ }\textbf {\bibinfo {volume} {88}},\ \bibinfo {pages} {319}
  (\bibinfo {year} {1997})}\BibitemShut {NoStop}%
\bibitem [{\citenamefont {Rajewski}\ \emph {et~al.}(1996)\citenamefont
  {Rajewski}, \citenamefont {Schadschneider},\ and\ \citenamefont
  {Schreckenberg}}]{RSS96}%
  \BibitemOpen
  \bibfield  {author} {\bibinfo {author} {\bibfnamefont {N.}~\bibnamefont
  {Rajewski}}, \bibinfo {author} {\bibfnamefont {A.}~\bibnamefont
  {Schadschneider}}, \ and\ \bibinfo {author} {\bibfnamefont {M.}~\bibnamefont
  {Schreckenberg}},\ }\href@noop {} {\bibfield  {journal} {\bibinfo  {journal}
  {J. Phys. A}\ }\textbf {\bibinfo {volume} {29}},\ \bibinfo {pages} {L305}
  (\bibinfo {year} {1996})}\BibitemShut {NoStop}%
\bibitem [{\citenamefont {Rajewski}\ and\ \citenamefont
  {Schreckenberg}(1997)}]{RS97}%
  \BibitemOpen
  \bibfield  {author} {\bibinfo {author} {\bibfnamefont {N.}~\bibnamefont
  {Rajewski}}\ and\ \bibinfo {author} {\bibfnamefont {M.}~\bibnamefont
  {Schreckenberg}},\ }\href@noop {} {\bibfield  {journal} {\bibinfo  {journal}
  {Physica A}\ }\textbf {\bibinfo {volume} {139}},\ \bibinfo {pages} {245}
  (\bibinfo {year} {1997})}\BibitemShut {NoStop}%
\bibitem [{\citenamefont {Evans}\ \emph {et~al.}(1998)\citenamefont {Evans},
  \citenamefont {Rajewsky},\ and\ \citenamefont {Speer}}]{ERS99}%
  \BibitemOpen
  \bibfield  {author} {\bibinfo {author} {\bibfnamefont {M.~R.}\ \bibnamefont
  {Evans}}, \bibinfo {author} {\bibfnamefont {N.}~\bibnamefont {Rajewsky}}, \
  and\ \bibinfo {author} {\bibfnamefont {E.~R.}\ \bibnamefont {Speer}},\
  }\href@noop {} {\bibfield  {journal} {\bibinfo  {journal} {J. Stat. Phys.}\
  }\textbf {\bibinfo {volume} {95}},\ \bibinfo {pages} {45} (\bibinfo {year}
  {1998})}\BibitemShut {NoStop}%
\bibitem [{\citenamefont {de~Gier}\ and\ \citenamefont
  {Nienhuis}(1999)}]{dGN99}%
  \BibitemOpen
  \bibfield  {author} {\bibinfo {author} {\bibfnamefont {J.}~\bibnamefont
  {de~Gier}}\ and\ \bibinfo {author} {\bibfnamefont {B.}~\bibnamefont
  {Nienhuis}},\ }\href@noop {} {\bibfield  {journal} {\bibinfo  {journal}
  {Phys. Rev. E}\ }\textbf {\bibinfo {volume} {59}},\ \bibinfo {pages} {4899}
  (\bibinfo {year} {1999})}\BibitemShut {NoStop}%
\bibitem [{\citenamefont {W\"{o}lki}(2005)}]{W05}%
  \BibitemOpen
  \bibfield  {author} {\bibinfo {author} {\bibfnamefont {M.}~\bibnamefont
  {W\"{o}lki}},\ }\href@noop {} {\emph {\bibinfo {title} {Steady States of
  Discrete Mass Transport Models (Master thesis)}}}\ (\bibinfo {address}
  {University of Duisburg-Essen},\ \bibinfo {year} {2005})\BibitemShut
  {NoStop}%
\bibitem [{\citenamefont {Derbyshev}\ \emph {et~al.}(2012)\citenamefont
  {Derbyshev}, \citenamefont {Poghosyan}, \citenamefont {Povolotsky},\ and\
  \citenamefont {Priezzhev}}]{DPPP}%
  \BibitemOpen
  \bibfield  {author} {\bibinfo {author} {\bibfnamefont {A.~E.}\ \bibnamefont
  {Derbyshev}}, \bibinfo {author} {\bibfnamefont {S.~S.}\ \bibnamefont
  {Poghosyan}}, \bibinfo {author} {\bibfnamefont {A.~M.}\ \bibnamefont
  {Povolotsky}}, \ and\ \bibinfo {author} {\bibfnamefont {V.~B.}\ \bibnamefont
  {Priezzhev}},\ }\href@noop {} {\bibfield  {journal} {\bibinfo  {journal} {J.
  Stat. Mech.}\ }\textbf {\bibinfo {volume} {2012}},\ \bibinfo {pages} {P05014}
  (\bibinfo {year} {2012})}\BibitemShut {NoStop}%
\bibitem [{\citenamefont {Derbyshev}\ \emph {et~al.}(2015)\citenamefont
  {Derbyshev}, \citenamefont {Povolotsky},\ and\ \citenamefont
  {Priezzhev}}]{DPP15}%
  \BibitemOpen
  \bibfield  {author} {\bibinfo {author} {\bibfnamefont {A.~E.}\ \bibnamefont
  {Derbyshev}}, \bibinfo {author} {\bibfnamefont {A.~M.}\ \bibnamefont
  {Povolotsky}}, \ and\ \bibinfo {author} {\bibfnamefont {V.~B.}\ \bibnamefont
  {Priezzhev}},\ }\href@noop {} {\bibfield  {journal} {\bibinfo  {journal}
  {Phys. Rev. E}\ }\textbf {\bibinfo {volume} {91}},\ \bibinfo {pages} {022125}
  (\bibinfo {year} {2015})}\BibitemShut {NoStop}%
\bibitem [{\citenamefont {Aneva}\ and\ \citenamefont {Brankov}(2016)}]{AB16}%
  \BibitemOpen
  \bibfield  {author} {\bibinfo {author} {\bibfnamefont {B.~L.}\ \bibnamefont
  {Aneva}}\ and\ \bibinfo {author} {\bibfnamefont {J.~G.}\ \bibnamefont
  {Brankov}},\ }\href@noop {} {\bibfield  {journal} {\bibinfo  {journal} {Phys.
  Rev. E}\ }\textbf {\bibinfo {volume} {94}},\ \bibinfo {pages} {022138}
  (\bibinfo {year} {2016})}\BibitemShut {NoStop}%
\bibitem [{\citenamefont {Bunzarova}\ and\ \citenamefont
  {Pesheva}(2017)}]{BunP17}%
  \BibitemOpen
  \bibfield  {author} {\bibinfo {author} {\bibfnamefont {N.~Z.}\ \bibnamefont
  {Bunzarova}}\ and\ \bibinfo {author} {\bibfnamefont {N.~C.}\ \bibnamefont
  {Pesheva}},\ }\href@noop {} {\bibfield  {journal} {\bibinfo  {journal} {Phys.
  Rev. E}\ }\textbf {\bibinfo {volume} {95}},\ \bibinfo {pages} {052105}
  (\bibinfo {year} {2017})}\BibitemShut {NoStop}%
\bibitem [{\citenamefont {Bunzarova}\ \emph {et~al.}(2017)\citenamefont
  {Bunzarova}, \citenamefont {Pesheva}, \citenamefont {Priezzhev},\ and\
  \citenamefont {Brankov}}]{BPPB17}%
  \BibitemOpen
  \bibfield  {author} {\bibinfo {author} {\bibfnamefont {N.~Z.}\ \bibnamefont
  {Bunzarova}}, \bibinfo {author} {\bibfnamefont {N.~C.}\ \bibnamefont
  {Pesheva}}, \bibinfo {author} {\bibfnamefont {V.}~\bibnamefont {Priezzhev}},
  \ and\ \bibinfo {author} {\bibfnamefont {J.~G.}\ \bibnamefont {Brankov}},\
  }\href@noop {} {\bibfield  {journal} {\bibinfo  {journal} {J. Phys: Conf.
  Series}\ }\textbf {\bibinfo {volume} {936}},\ \bibinfo {pages} {012026}
  (\bibinfo {year} {2017})}\BibitemShut {NoStop}%
\bibitem [{\citenamefont {Brankov}\ \emph {et~al.}(2018)\citenamefont
  {Brankov}, \citenamefont {Bunzarova}, \citenamefont {Pesheva},\ and\
  \citenamefont {Priezzhev}}]{BBPP18}%
  \BibitemOpen
  \bibfield  {author} {\bibinfo {author} {\bibfnamefont {J.~G.}\ \bibnamefont
  {Brankov}}, \bibinfo {author} {\bibfnamefont {N.~Z.}\ \bibnamefont
  {Bunzarova}}, \bibinfo {author} {\bibfnamefont {N.~C.}\ \bibnamefont
  {Pesheva}}, \ and\ \bibinfo {author} {\bibfnamefont {V.}~\bibnamefont
  {Priezzhev}},\ }\href@noop {} {\bibfield  {journal} {\bibinfo  {journal}
  {Physica A}\ }\textbf {\bibinfo {volume} {494}},\ \bibinfo {pages} {340}
  (\bibinfo {year} {2018})}\BibitemShut {NoStop}%
\bibitem [{\citenamefont {Hrab{\'a}k}(2014)}]{PhD}%
  \BibitemOpen
  \bibfield  {author} {\bibinfo {author} {\bibfnamefont {P.}~\bibnamefont
  {Hrab{\'a}k}},\ }\href@noop {} {\emph {\bibinfo {title} {Ph.D. Thesis}}}\
  (\bibinfo  {publisher} {Czech Technical University, Faculty of Nuclear
  Sciences and Physical Engineering},\ \bibinfo {address} {Praga},\ \bibinfo
  {year} {2014})\BibitemShut {NoStop}%
\end{thebibliography}%

\end{document}